\documentclass[aps,prb,twocolumn,showpacs]{revtex4-1}
\usepackage[latin1]{inputenc}
\usepackage{amsmath}
\usepackage[T1]{fontenc}
\usepackage{graphicx}
\usepackage{epstopdf}
\usepackage[LGRgreek]{mathastext}
\usepackage{graphicx}
\usepackage{grffile}
\usepackage{textcomp}
\usepackage{gensymb}
\usepackage{longtable}
\usepackage{lipsum}
\usepackage{array}

\begin{document}
\title{Effect of disorder on the quantum spin liquid candidate Na$_4$Ir$_3$O$_8$} 
\author{Ashiwini Balodhi and Yogesh Singh}
\affiliation{Indian Institute of Science Education and Research (IISER) Mohali,
Knowledge City, Sector 81, Mohali 140306, India}

\begin{abstract}
We report on the effects of introducing magnetic and non-magnetic  disorder in the hyperkagome iridate quantum spin liquid (QSL) candidate Na$_4$Ir$_3$O$_8$ by partially replacing Ir$^{4+}$ ($S = 1/2$) with Ru$^{4+}$ ($S = 1$) or Ti$^{4+}$ ($S = 0$).  Specifically, we synthesized Na$_4$(Ir$_{1-x}$Ru$_x$)$_3$O$_8 (x = 0.05, 0.10, 0.2, 0.3)$ and Na$_4$Ir$_{2.7}$Ti$_{0.3}$O$_8$ samples and measured electrical transport, AC and DC magnetization, and heat capacity down to $T = 1.8$ K. Na$_4$Ir$_3$O$_8$ is associated with a large Weiss temperature $\theta = -650$ K, a broad anomaly in magnetic heat capacity C$_{mag}$ at T $\approx25$~K, low temperature power-law heat capacity, and spin glass freezing below $T_f \approx 6$~K\@.  We track the change in these characteristic features as Ir is partially substituted by Ru or Ti.  We find that for Ru substitution, $\theta$ increases and stays negative, the anomaly in C$_{mag}$ is suppressed in magnitude and pushed to lower temperatures, low temperature $C \sim T^\alpha$ with $\alpha$ between $2$ and $3$ and decreasing towards $2$ with increasing $x$, and $T_f$ increases with increase in Ru concentration $x$. For Ti substitution we find that $\theta$ and T$_f$ become smaller and the anomaly in $C_{mag}$ is completely suppressed.  In addition, introducing non-magnetic Ti leads to the creation of orphan spins which show up in the low temperature magnetic susceptibility.

\end{abstract}
\maketitle
\section{Introduction}

The study of 4-d and 5-d transition metal oxides (TMOs) has become one of the active areas of research in condensed matter physics due to the interplay between various energy scales like crystalline electric field, electronic correlations, Hund's coupling, spin-orbit coupling, and lattice distortions \cite{rau}. The cooperative effects of these different energy scales has led to several novel behaviours like the $J_{eff}= 1/2$ Mott insulating state in  Sr$_2$IrO$_4$  \cite{kim,Cao}, topological Mott insulators  \cite{balent}, and Kitaev materials Na$_2$IrO$_3$, $\alpha$-RuCl$_3$~ \cite{Winter2017,J}, to name a few. 

The 5-d TMO Na$_4$Ir$_3$O$_8$ is a novel material which combines high spin-orbit coupling, possibly $J_{eff} = 1/2$ moments with a geometrically frustrated hyperkagome lattice \cite{okamato, yogesh}.  The lattice topology coupled with the low spin made this material an exciting candidate for the observation of 3-dimensional (3D) quantum spin liquid (QSL) state.  Indeed first experiments showed $S = 1/2$ Ir$^{4+}$ moments on a highly frustrated hyper-kagome lattice.  This resulted in the absence of long ranged magnetic order inspite of large antiferromagnetic magnetic exchange as evidenced by a large Weiss temperature $\theta = -650$~K~ \cite{okamato}.  Additionally, a power-law heat capacity at low temperatures and a broad anomaly in the magnetic heat capacity $C_{mag}$ strongly suggested the realization of a 3D gapless QSL state in Na$_4$Ir$_3$O$_8$ \cite{okamato}.  Several theoretical studies supported a QSL state although details of the state varied in different approaches \cite{lawer,chen1,chen2,zhou, Podolsky, Norman}. Prediction of a gapless QSL having a spinon Fermi surface with line nodes were also made \cite{lawer,zhou}.  Unfortunately a spin-glassy state was observed giving rise to magnetic irreversibility at low temperatures ($\sim 6$~K) \cite{okamato, yogesh}.   This frozen state was confirmed by microscopic measurements like muon spin rotation ($\mu$SR) and neutron diffraction measurements \cite{Dally2014} and $^{23}$Na and $^{17}$O NMR measurements \cite{shockley}.  Recently, the possibility of anisotropic Kitaev interactions were also discussed theoretically for Na$_4$Ir$_3$O$_8$~\cite{Kimchi2014}.  Experimental Raman scattering response is also consistent with strong Kitaev interactions with smaller Heisenberg terms \cite{gupta}.

Recently we have reported on a study of Na deficient materials Na$_{4-x}$Ir$_3$O$_8$ and found the spin-liquid like features mentioned previously to be robust against large Na removal \cite{ashiwini}.  In that study, the Ir magnetic sub-lattice was not disturbed.  In the current work we present a comprehensive study of the electrical transport, magnetic, and thermal properties of Na$_4$Ir$_3$O$_8$ materials in which magnetic or non-magnetic disorder is introduced into the magnetic sub-lattice by partially replacing Ir by Ru/Ti.  In particular, we synthesized a series of polycrystalline samples Na$_4$(Ir$_{1-x}$Ru$_x$)$_3$O$_8 (x = 0.05, 0.10, 0.2, 0.3)$ and Na$_4$Ir$_{2.7}$Ti$_{0.3}$O$_8$.  We track the change in Curie-Weiss temperature $\theta$, in the magnitude and temperature dependence of the anomaly in C$_{mag}$, the exponent of the low temperature power-law dependence of heat capacity, and the spin freezing temperature $T_f$ with increasing Ru/Ti concentration $x$.  

\section{Experimental details}

Polycrystalline samples of Na$_4$(Ir$_{1-x}$Ru$_x$)$_3$O$_8$  (x= 0, 0.05, 0.1, 0.2 and 0.3) and Na$_4$Ir$_{2.7}$Ti$_{0.3}$O$_8$ were synthesized from Na$_2$CO$_3$, Ir and Ru/Ti (99.99 \%- Alpha-Aesar) metal powders by conventional solid state reaction.  Stoichiometric amounts of the high-purity starting ingredients were mixed thoroughly and loaded into alumina crucibles with a lid. These mixtures were first calcined at 750 $^o$C for 22 hrs.  The resulting mixtures were ground and pressed into a pellet (10 mm) and subsequently heated twice at 975$^o$C for 35 hrs in oxygen with an intermediate grinding and pelletizing step. After the final heating, the samples were quenched in air at $750~^o$C\@.  The Powder X-Ray diffraction (PXRD) was collected at room temperature with a Rikagu diffractometer (Cu K$\alpha$) and analyzed by GSAS using Rietveld refinement. Electrical transport, DC magnetization, AC susceptibility, and heat capacity measurements were done using a Quantum Design Physical Property Measurement System (QD-PPMS).  All materials were found to show an insulating temperature dependence of their electrical resistance.

\begin{figure}[htp]
	\includegraphics[width=9 cm]{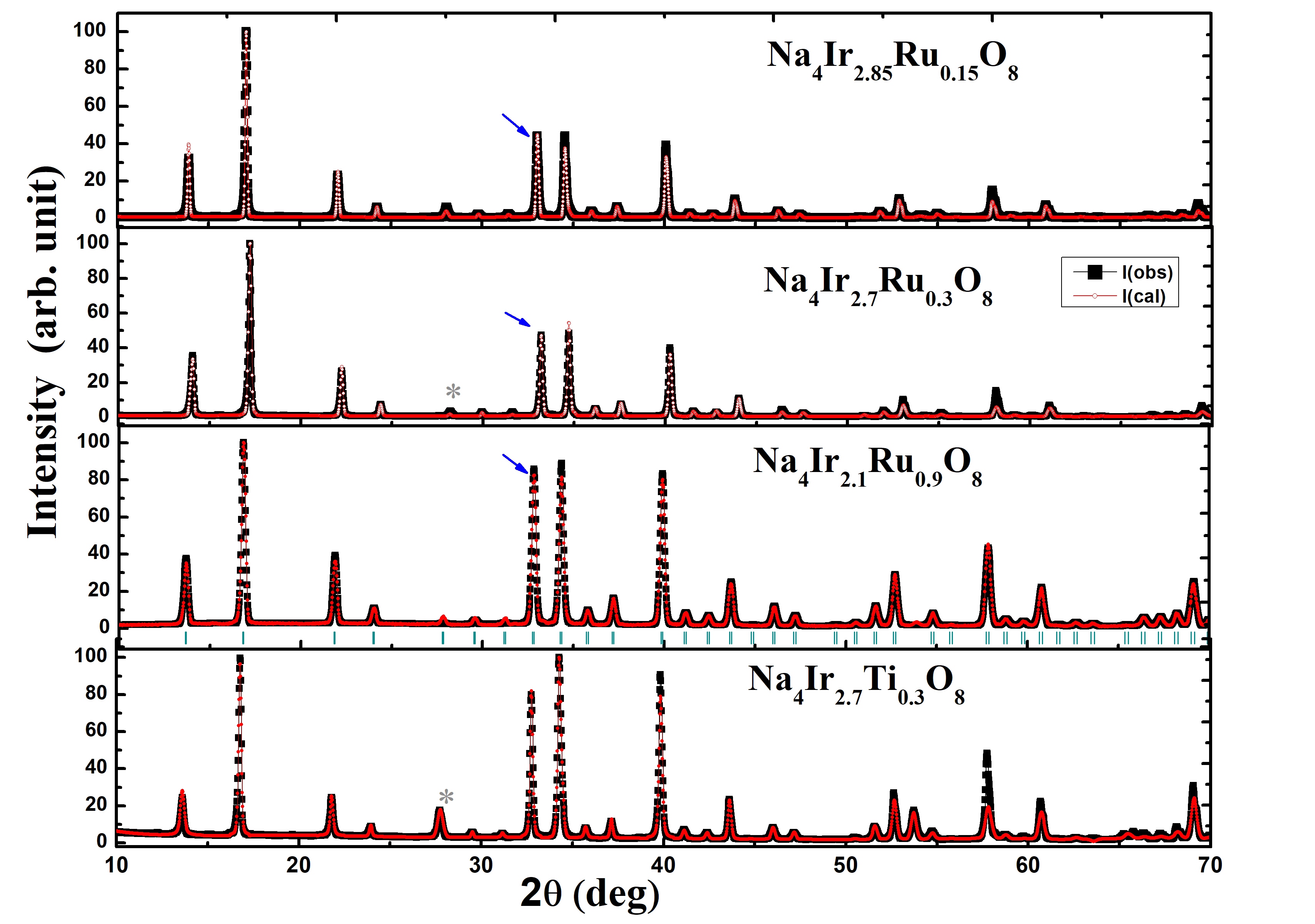}
	\caption{(Color online) Powder x-ray diffraction pattern for Na$_4$(Ir$_{1-x}$Ru$_x$)$_3$O$_8$ (x = 0.05, 0.1, 0.3) and Na$_4$Ir$_{2.7}$Ti$_{0.3}$O$_8$ materials (open symbols). The solid curve through the data is the Rietveld refinement. The $\ast$ marks the position of the largest diffraction peak for the IrO$_2$ impurity phase found in the samples. The arrow marks the position of the (311) reflection whose relative intensity varies the most.}
	 \label{xrd1}
\end{figure}

\begin{table*}[t]
	\centering
	\caption{ Structural parameters for Na$_4$(Ir$_{1-x}$Ru$_x$)$_3$O$_8$ (x = 0.5, 0.1, 0.2 and 0.3) and Na$_4$Ir$_{2.7}$Ti$_{0.3}$O$_8$ obtained from a Rietveld refinement of room temperature powder x-ray patterns shown in Fig.~\ref{xrd1} with space group 213, P4$_1$32.}
	
	\begin{tabular}  {|ccccccc| }	
		\hline \hline	
		Atom & Wyck & x & y& z& Occ. & B (\AA)  \\ 
		\hline
Na$_4$Ir$_{2.85}$Ru$_{0.15}$O$_8$ & a = 8.9848(5)~\AA &    Cell volume = 725.324(11) \AA$^3$       & IrO$_2$ $\approx$ 3 \% & & & \\ \hline
		Ir   &12d & 1/8        &0.141       & 0.392     &0.94(1)    & 0.009    \\
		Ru   & 12d & 1/8        &0.141       & 0.392      &0.06(2)     & 0.0024  \\
		Na1  &4a  & 3/8        & 3/8           & 3/8           &1         & 0.009   \\
		Na2  &4b  & 7/8        & 7/8           & 7/8           &0.72(1)   & 0.008  \\
		Na3  &12d & 0.125      & 0.1564        & 0.4064        &0.69(3)      & 0.0042  \\ 
		O1   &8c  & 0.6181     & x             & x             &1         & 0.006   \\
		O2   &24e &0.1368      &0.8287         &0.8592         &1         & 0.025  \\ \hline

		\\ Na$_4$Ir$_{2.7}$Ru$_{0.3}$O$_8$ & a = 8.9839(4)~\AA &     Cell volume = 725.096(16) \AA$^3$   & IrO$_2$ $\approx$ 4 \% \\ \hline
		Ir   & 12d & 1/8        &0.8825      & 0.1325       & 0.89(3)      & 0.0039  \\
		Ru   & 12d & 1/8        &0.8825      & 0.1325       & 0.11(3)      & 0.0026  \\
		Na1  &4a  & 3/8        & 3/8        & 3/8        &1         & 0.0069  \\
		Na2  &4b  & 7/8        & 7/8        & 7/8        &0.73(2)      & 0.0009 \\
		Na3  &12d & 0.125      &0.12505     & 0.3106     &0.64(3)      & 0.0028  \\
		O1   &8c  & 0.6654     &x           & x          &1         & 0.0046  \\ 
		O2   &24e & 0.1467     & 0.9809     & 0.9068     &1         & 0.025   \\ \hline

		\\ Na$_4$Ir$_{2.4}$Ru$_{0.6}$O$_8$ &       a =  8.9827(5)  \AA &       Cell volume = 724.80(7) \AA$^3$    & IrO$_2$ $\approx$ 5 \% \\ \hline
		Ir   &12d 		& 1/8        &0.8854      & 0.135     &0.80(2)       & 0.009  \\ 
		Ru   &12d 		& 1/8        &0.8854      & 0.135     & 0.20(3)      & 0.0009  \\
		Na1  &4a  		& 3/8        & 3/8        & 3/8        &1        & 0.0025  \\ 
		Na2  &4b  		& 7/8        & 7/8        & 7/8        &0.70(1)     & 0.0028  \\
		Na3  &12d 		& 0.125       &0.141900      & 0.391900     &0.65(1)     & 0.0025 \\ 
		O1   &8c  		& 0.629706   & x          & x          &1       & 0.0025  \\ 
		O2   &24e 		& 0.1439     & 0.8956     & 0.9025     & 1      & 0.0025  \\ \hline	
			
\\ Na$_4$Ir$_{2.1}$Ru$_{0.9}$O$_8$ & a = 8.9812(3)~\AA &     Cell volume = 724.4505(9) \AA$^3$ &      & IrO$_2$ $\approx$ 5 \% \\ \hline
		Ir   &12d & 1/8        &0.9025      & 0.1588     &0.70(1)       & 00013  \\ 
		Ru   &12d & 1/8        &0.9025      & 0.1588     & 0.30(1)      & 0.0074  \\
		Na1  &4a  & 3/8        & 3/8        & 3/8        &1        & 0.0025  \\ 
		Na2  &4b  & 7/8        & 7/8        & 7/8        &0.68(2)     & 0.0028  \\
		Na3  &12d & 0.125       &0.1544      & 0.4044     &0.61(2)     & 0.0025  \\ 
		O1   &8c  & 0.6339     & x          & x          &1       & 0.0025  \\ 
		O2   &24e & 0.1354     & 0.8972     & 0.9007     & 1      & 0.0025  \\ \hline
		
		\\ Na$_4$Ir$_{2.7}$Ti$_{0.3}$O$_8$ & a = 8.9684(7)~\AA &  Cell volume = 721.349(22)  \AA$^3$    & IrO$_2$ $\approx$ 7 \% \\ \hline
		Ir  	 &12d 		& 1/8       &0.8828     & 0.1328     &0.89(2)       & 0.002  \\ 
		Ti  	 &12d 		& 1/8       &0.8828      & 0.1328     & 0.11(1)      & 0.009  \\
		Na1	  &4a  & 3/8        & 3/8        & 3/8        &1        & 0.0025  \\ 
		Na2	  &4b  & 7/8        & 7/8        & 7/8        &0.70(2)     & 0.0013  \\
		Na3 	 &12d & 0.125       &0.1575      & 0.4075     &0.65(2)     & 0.0078  \\ 
		O1  	 & 8c  & 0.6316     & x          & x          &1       & 0.0047  \\ 
		O2   	&24e & 0.1399     & 0.8760     & 0.9052     & 1      & 0.0025  \\ \hline	
	\end{tabular}
	\label{RuTable1}
	
\end{table*}

\begin{figure}[htp]
	\includegraphics[width= 8 cm]{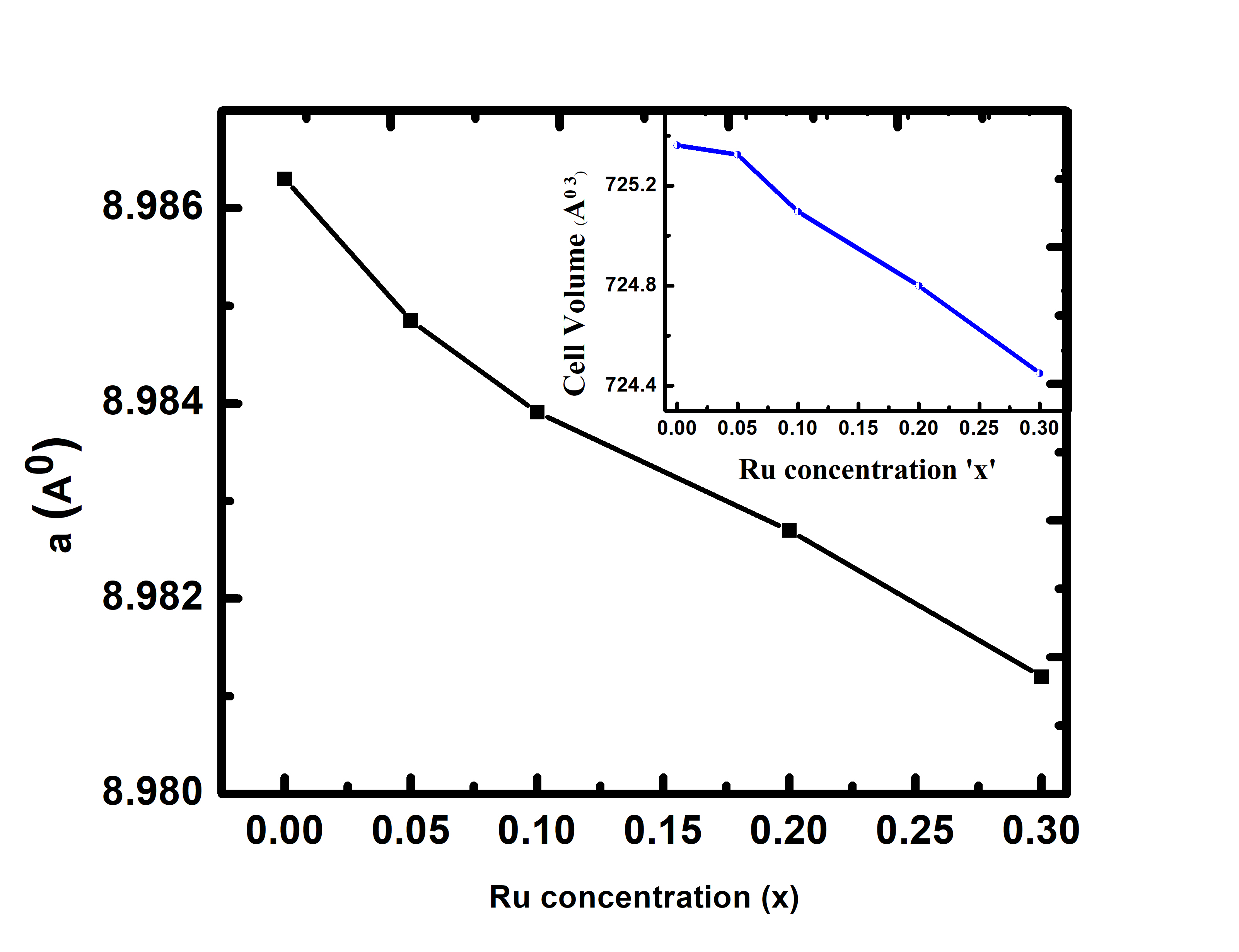}
	\caption{Variation of lattice parameter '$a$' with Ru concentration for Na$_4$(Ir$_{1-x}$Ru$_x$)$_3$O$_8~ (x = 0, 0.05, 0.10, 0.2, 0.3)$.  The inset shows the variation of the unit cell volume.}
	\label{xrd2}
\end{figure}

\section{Results} 
\subsection{Structure}
Room temperature PXRD for Na$_4$(Ir$_{1-x}$Ru$_x$)$_3$O$_8$ with different Ru concentrations x = 0.05, 0.10, 0.2, 0.3 and Na$_4$Ir$_{2.7}$Ti$_{0.3}$O$_8$ along with a Rietveld refinement of the data are shown in Fig.~\ref{xrd1}.  All samples show the expected PXRD pattern apart from a small trace of IrO$_2$ ($\sim 5 \%$) found in all samples which is marked by the $\ast$ in Fig.~\ref{xrd1} and is similar to recent reports \cite{okamato, ashiwini,Zheng}. Synthesis of samples with higher concentration of Ru/Ti substitution was attempted but single phase materials were not obtained.  Parameters obtained from a Rietveld refinement of the successfully synthesized materials are shown in Table~\ref{RuTable1}.  The variation of the cubic lattice constant `$a$' obtained from the refinements and the unit cell volume as a function of Ru concentration $x$ are shown in  Fig.~\ref{xrd2}.

Our structural analysis indicates a monotonic reduction of the cubic lattice parameter $a$ with increasing Ru concentration '$x$' which is expected from a comparison of the ionic size of Ir$^{4+}$  (0.625 \AA) to Ru$^{4+}$  (0.62 \AA). Similarly for Na$_4$Ir$_{2.7}$Ti$_{0.3}$O$_8$, the lattice parameter shrinks to $8.9684$~\AA ~(ionic size for Ti$^{4+} = 0.605$~\AA), which is $\approx 1\%$ reduction in the cell volume compared to the parent compound.  

The refined fractional occupation of Ir and Ru/Ti shown in the Table~\ref{RuTable1} matches well with the target concentration $x$ for all samples. This indicates that partial replacement of Ru/Ti for Ir in Na$_4$Ir$_3$O$_8$ was successful. During refinement it was observed that there was no appreciable change in the fractional occupation for the Na1 site. It was therefore fixed to $1$.  The major change occurs for the occupation of Na2 and Na3 sites.  This is consistent with our observation that the Bragg peaks (311) and (222) in the powder diffraction patterns to which the Na2 and Na3 atoms contribute the most, are the ones whose intensities are most affected by the Ru/Ti substitution.  We can conclude that the Ru/Ti substituted samples are Na deficient to varying but small degrees.  We already know what to expect for Na deficient Na$_4$Ir$_3$O$_8$~\cite{ashiwini} and therefore changes due to Ru/Ti substitutions can be separated from changes due to Na deficiency. Recently deficiency of Na was also found in high quality single crystals of Na$_4$Ir$_3$O$_8$ with no significant change to physical properties \cite{Zheng}. 

\subsection{Magnetic Properties}
\begin{figure}[htp]
\includegraphics[width= 8 cm]{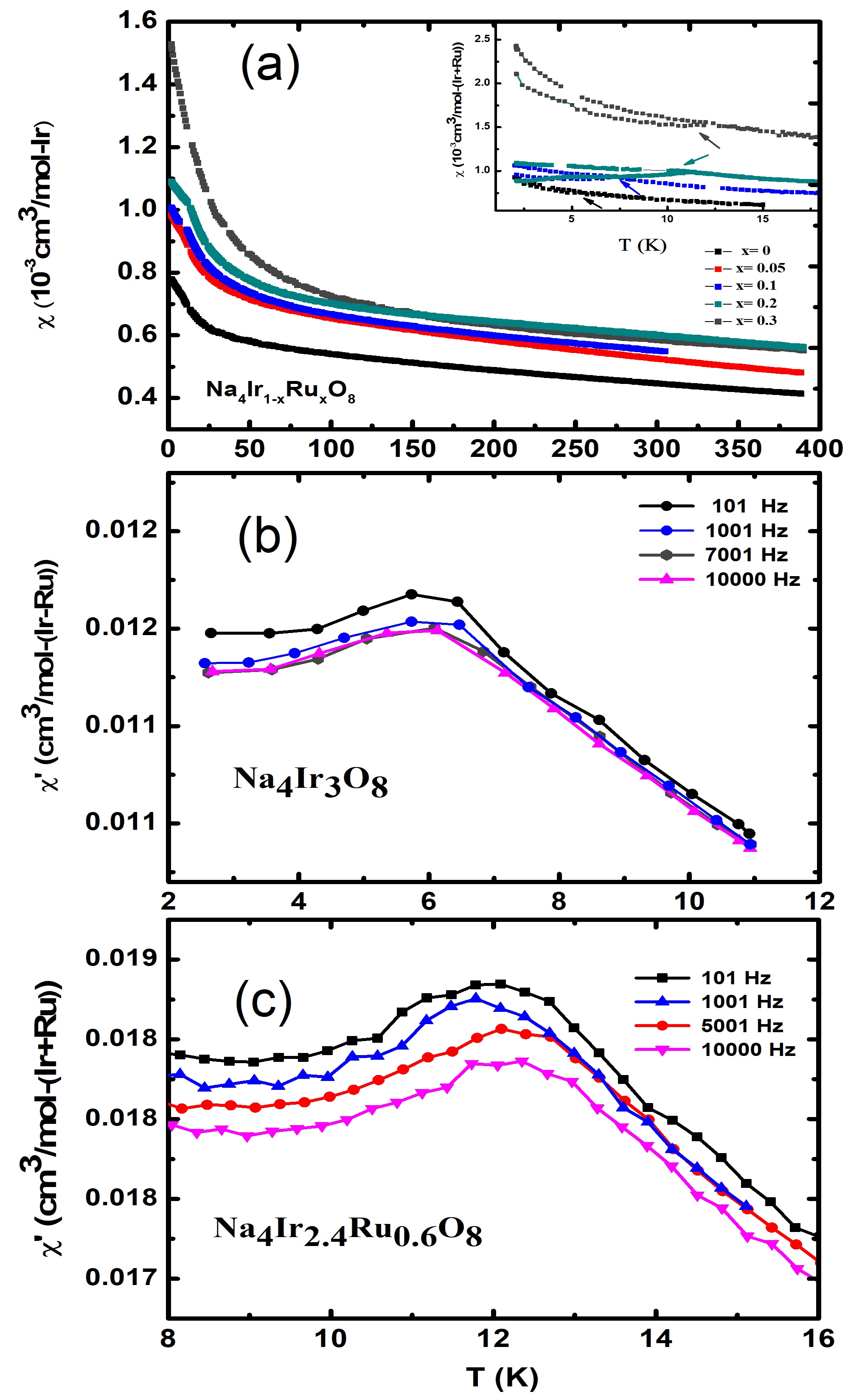}
\caption{ (color line) (a) DC Magnetic susceptibility $\chi$ versus temperature T(K) for Na$_4$(Ir$_{1-x}$Ru$_x)_3$O$_8$ (x = 0,0.05, 0.1, 0.2, 0.3) measured at a magnetic field $H = 1$~T\@. The solid curve through the data at high-temperatures for each $x$ is a fit by the Curie-Weiss expression in the temperature range $150$--$400$~K\@. Inset shows the ZFC-FC $\chi$ versus $T$ measured at $H = 200$~Oe for $x = 0, 0.05, 0.1, 0.2$ samples.  (b)  Real part of the AC magnetic susceptibility $\chi'$ versus $T$ for Na$_4$Ir$_3$O$_8$ and (c) Na$_4$Ir$_{2.4}$Ru$_{0.6}$O$_8$ measured with various excitation frequencies.} 
\label{fig-chi1}
\end{figure}

The temperature dependence of the dc magnetic susceptibitily $\chi = M/H$ of Na$_4$(Ir$_{1-x}$Ru$_x$)$_3$O$_8~ (x = 0.05, 0.10, 0.2, 0.3)$ in a magnetic field $H = 1$~T is shown in Fig.~\ref{fig-chi1}~(a). The $\chi$ for Na$_4$(Ir$_{1-x}$Ru$_x$)$_3$O$_8$ continually increases with increase in Ru concentration as expected on partially substituting Ir ($S = 1/2$) with Ru ($S = 1$) moments. The high-temperature data for all $x$ were found to follow a Curie-Weiss behaviour. The $\chi$(T) data for $T \geq 150$~K were fit by the Curie-Weiss expression $\chi = \chi_0 + {C\over T-\theta}$ where $\chi_0$, C, and $\theta$ are fitting parameters. The parameters obtained from the fit to the data for each $x$ are given in Table~\ref{Table-IV}. The value of the effective paramagnetic moment $\mu_{eff}$ for $x > 0$ is found to be larger than expected for $S =1/2$ from the Curie constant $C$.  This is consistent with expectations of substituting $S = 1/2$ with $S = 1$ ions.  The other thing to note is that $\theta$ continues to be large (it's magnitude actually increases with $x$) and negative indicating that strong antiferromagnetic interactions persist in all the Ru substituted samples. 

\begin{table}[htp]
\caption{Parameters obtained by an analysis of the magnetic susceptibility of Na$_4$(Ir$_{1-x}$Ru$_x$)$_3$O$_8 (x = 0,0.05, 0.1, 0.2, 0.3)$.}
 	\begin{tabular}  {|c|c|c|c|c|c|c|}	
 		\hline	
 		& $\chi_0 \frac{ 10^{-5} cm^3}{(Ir+Ru) mol}$ & $\theta$ (K) & C $\frac{cm^3K}{(Ir+Ru)mol}$  & $\mu_{eff}(\mu_B$)& $T_f$(K) \\ \hline 
 		x =   0    & 8.58 &-650 & 0.39 & 1.77& 6.3\\ \hline 
 		x = 0.05 & 2.006 & -689.4 & 0.48 & 1.96& 7.45 \\ \hline
 		x = 0.1  & 6.299 & -787.15 & 0.52& 2.03& 9.8\\ \hline 
 		x = 0.2  &  6.101  & -889  & 0.59 & 2.17& 11.6\\ \hline 
 		x = 0.3  &6.35 & -983  & 0.67 & 2.31& 13.2 \\ \hline 
 	\end{tabular}
 	\label{Table-IV}
 \end{table} 

Another feature of interest in the parent compound Na$_4$Ir$_3$O$_8$ is the spin freezing below $T_f\approx 6$~K\@.  To track the evolution of $T_f$ with $x$ the inset of Fig.~\ref{fig-chi1}~(a) shows the zero-field-cooled (ZFC) and field-cooled (FC) magnetization versus temperature data for various $x$ measured at $200$~Oe. The bifurcation between ZFC-FC marked by arrows in Fig.~\ref{fig-chi1}~(a) inset is taken as the $T_f$ and is seen clearly in all Ru substituted samples. It was found that $T_f$ increases significantly with $x$ reaching a value $T_f = 13.2$~K for $x = 0.3$.  The ZFC-FC bifurcation being a signature of a spin-glassy state is supported by our AC susceptibility $\chi_{ac}$ measurements.  

Figures~\ref{fig-chi1}~(b) and (c) show the frequency dependence of $\chi_{ac}$ versus $T$ for the $x = 0$ and $x = 0.3$ samples.  A clear maximum is observed in $\chi_{ac}$ at a temperature which matches with the bifurcation in ZFC-FC dc $\chi$ data.  The maximum, which moves to higher temperatures on increasing the excitation frequency is a strong signature of a glassy state \cite{Mydosh}.  Thus, the increased disorder due to Ru substitution results in a higher freezing temperature.
  
 \begin{figure}[htp]
 	\includegraphics[width= 8 cm]{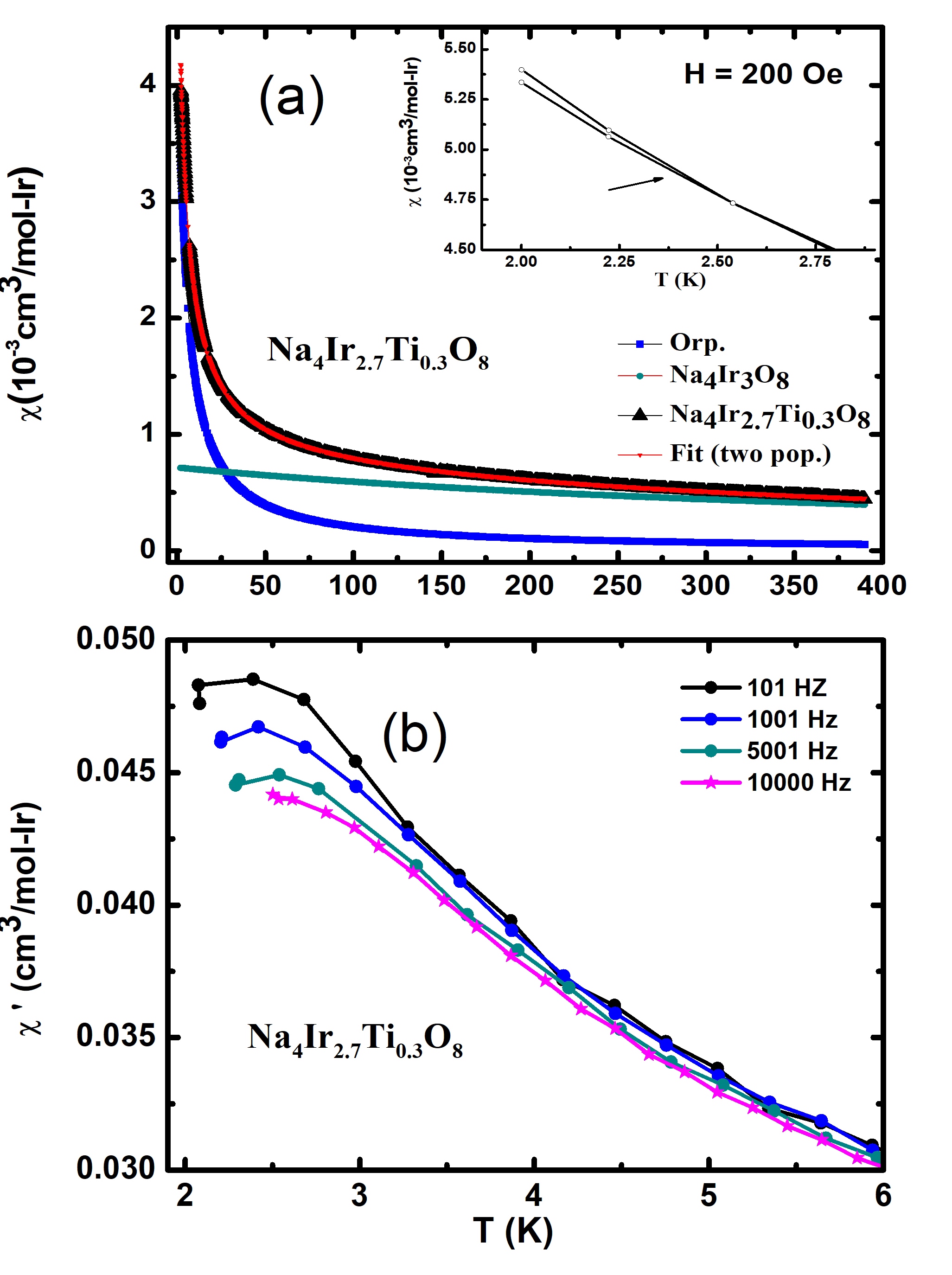}
 	\caption{(color line) DC Magnetic susceptibility $\chi$ versus $T$ for Na$_4$Ir$_{2.7}$Ti$_{0.3}$O$_8$.  A two population model fit is shown as the solid curve through the data.  The orphan spin contribution is also shown (see text for details).  Inset shows the ZFC-FC $\chi$ versus $T$ measured at $H = 200$~Oe.  (b) Real part of the AC Magnetic susceptibility $\chi'$ versus $T$ for Na$_4$Ir$_{2.7}$Ti$_{0.3}$O$_8$ measured with different excitation frequencies.}	
 	\label{fig-Ti}
 \end{figure}
 
A different picture emerges for non-magnetic Ti substitution.  Figure~\ref{fig-Ti}~(a) shows the $\chi$ versus $T$ for Na$_4$Ir$_{2.7}$Ti$_{0.3}$O$_8$.  We observe that non-magnetic Ti$^{4+}$ substitution results in an enhanced paramagnetic susceptibility.  This is most likely due to the creation of orphan spins \cite{Schiffer} i.e. some of the Ir spins behave as free spins and contribute an extra Curie paramagnetic term. We therefore used a phenomenological two-population model \cite{Schiffer} to analyze the $\chi(T)$ data for Na$_4$Ir$_{2.7}$Ti$_{0.3}$O$_8$.  The data were fit by the expression $\chi = \chi_0 + \frac{C_1}{(T+\theta_1)} + \frac{C_2}{(T+\theta_2)}$, where $C_1$ and $\theta_1$, and $C_2$ and $\theta_2$ are the Curie constant and Curie-Weiss temperature for the correlated and orphan spins, respectively \cite{Schiffer}. The fit, shown as the solid curve through the data in Fig.~\ref{fig-Ti}, gave the values $C_1 = 0.36$~cm$^3$-K/mol-(Ir+Ru), $C_2= 0.017$~ cm$^3$-K/mol-(Ir+Ru), $\theta_1 = -500$~K, and $\theta_2 = -4.5$~K\@.  The Curie constant for orphan spins $C_2= 0.017$~ cm$^3$-K/mol-(Ir+Ru) is equivalent to $\approx 5~\%$ of $S = 1/2$.  Additionally, a significant suppression in the magnetic energy scale $\theta_1$ is observed. The ZFC-FC magnetization versus $T$ for Na$_4$Ir$_{2.7}$Ti$_{0.3}$O$_8$ is shown in the inset of Fig-\ref{fig-Ti}~(a). From these data we conclude that the freezing temperature is suppressed down to $T_f \approx 2.3$~K\@.  Thus non-magnetic substitution has the opposite effect on $T_f$ compared to magnetic substitution. 

The $\chi_{ac}(T)$ data for Na$_4$Ir$_{2.7}$Ti$_{0.3}$O$_8$ measured at various excitation frequencies is shown in Fig.~\ref{fig-Ti}~(b).  A clear maximum is observed near $2.3$~K at low frequency and this maximum moves to higher temperatures for higher frequencies, a clear signature of spin-glassy behaviour below $T_f \approx 2.3$~K ~\cite{Mydosh}. 

\subsection{Heat Capacity} 
 \begin{figure}
 	\centering
 	\includegraphics[width= 6 cm]{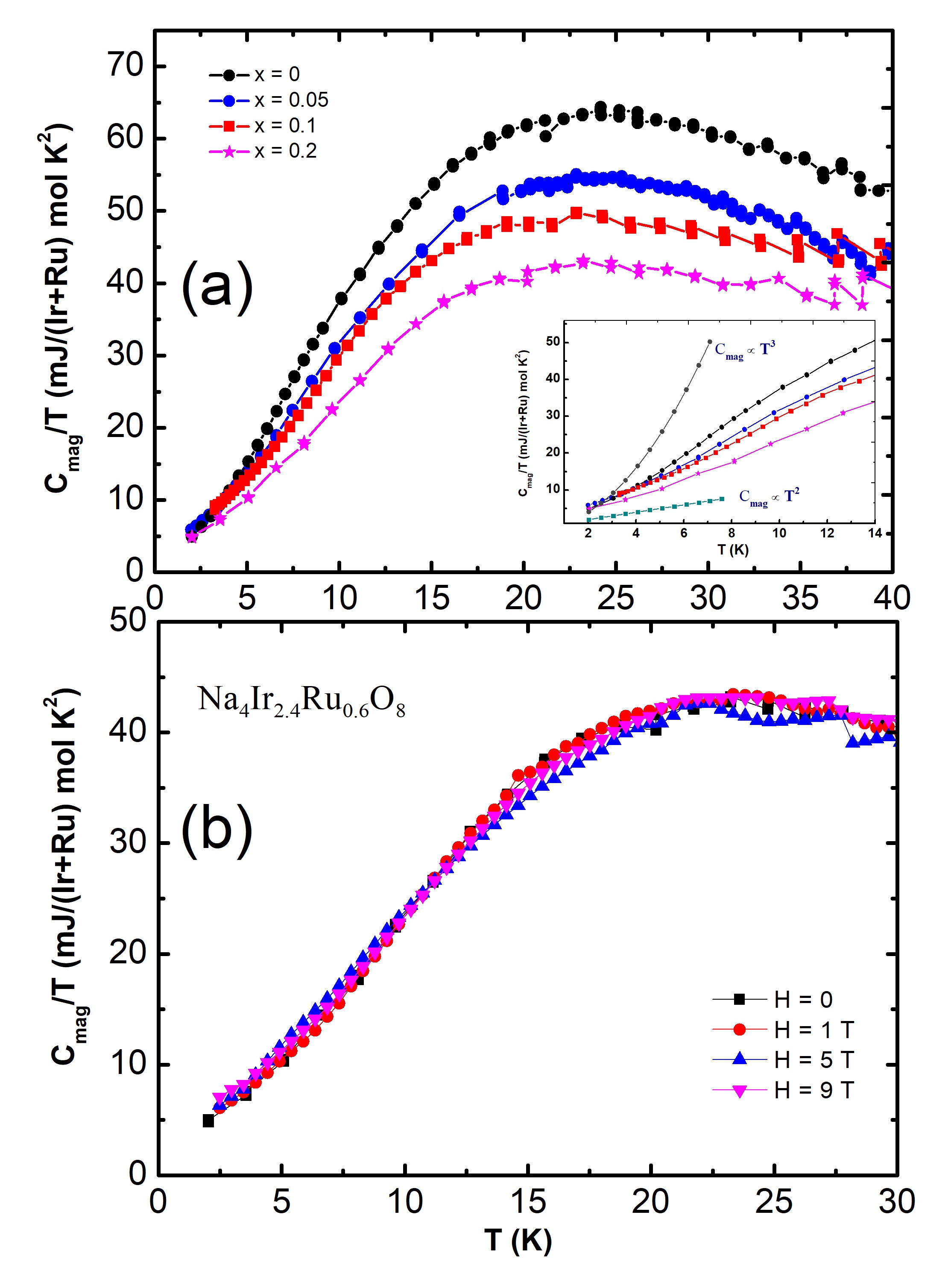}
 	\caption{ (a) $C_{mag}/T$ versus $T$ for Na$_4$(Ir$_{1-x}$Ru$_x$)$_3$O$_8~ (x = 0, 0.1, 0.05, 0.2)$. Inset shows the low temperature power-law $T$ dependence of $C_{mag}/T$ for Na$_4$(Ir$_{1-x}$Ru$_x$)$_3$O$_8$.  (b) $C_{mag}/T$ versus $T$ for Na$_4$Ir$_{2.4}$Ru$_{0.6}$O$_8$ at different applied magnetic fields.}
 	\label{hc1}
 \end{figure}
 
The magnetic contribution to the heat capacity $C_{mag}$ versus $T$ data for the Ru substituted samples Na$_4$(Ir$_{1-x}$Ru$_x$)$_3$O$_8~ (x = 0, 0.05, 0.1, 0.2)$ are plotted in Fig.~\ref{hc1}~(a). The heat capacity of the iso-structural non-magnetic material Na$_4$Sn$_3$O$_8$ was used to estimate the lattice contribution to the heat capacity for Na$_4$(Ir$_{1-x}$Ru$_x$)$_3$O$_8$.  The parent compound Na$_4$Ir$_3$O$_8$ is known to show a broad anomaly in magnetic heat capacity around $T \sim 25$~K\@.  As is evident from Fig.~\ref{hc1}~(a) this broad anomaly in $C_{mag}$ is suppressed in magnitude, broadens, and the maximum is shifted to lower temperatures with increasing $x$. 
The inset in Fig.~\ref{hc1}~(a) shows the variation of $C_{mag}/T$ versus $T$ at low temperatures to highlight the power-law behaviour of $C_{mag}$. These data show power law $C \sim T^\alpha$ behaviour with an exponent $\alpha$ between $2$ and $3$ as has been reported previously for the $x = 0$ material \cite{okamato, yogesh}.  The exponent was found to decrease with increasing $x$ and reaches $\approx 2$ for $x = 0.2$ (not shown). 

Figure~\ref{hc1}~(b) shows the $C_{mag}/T$ vs $T$ data for Na$_4$Ir$_{2.4}$Ru$_{0.6}$O$_8$ measured at various applied magnetic fields.  It was found that the broad anomaly in $C_{mag}$ and the exponent of the low temperature power-law behaviour are largely insensitive to magnetic fields up to $H = 9$~T\@.
 
  \begin{figure}
 	\centering
	\includegraphics[width= 6 cm]{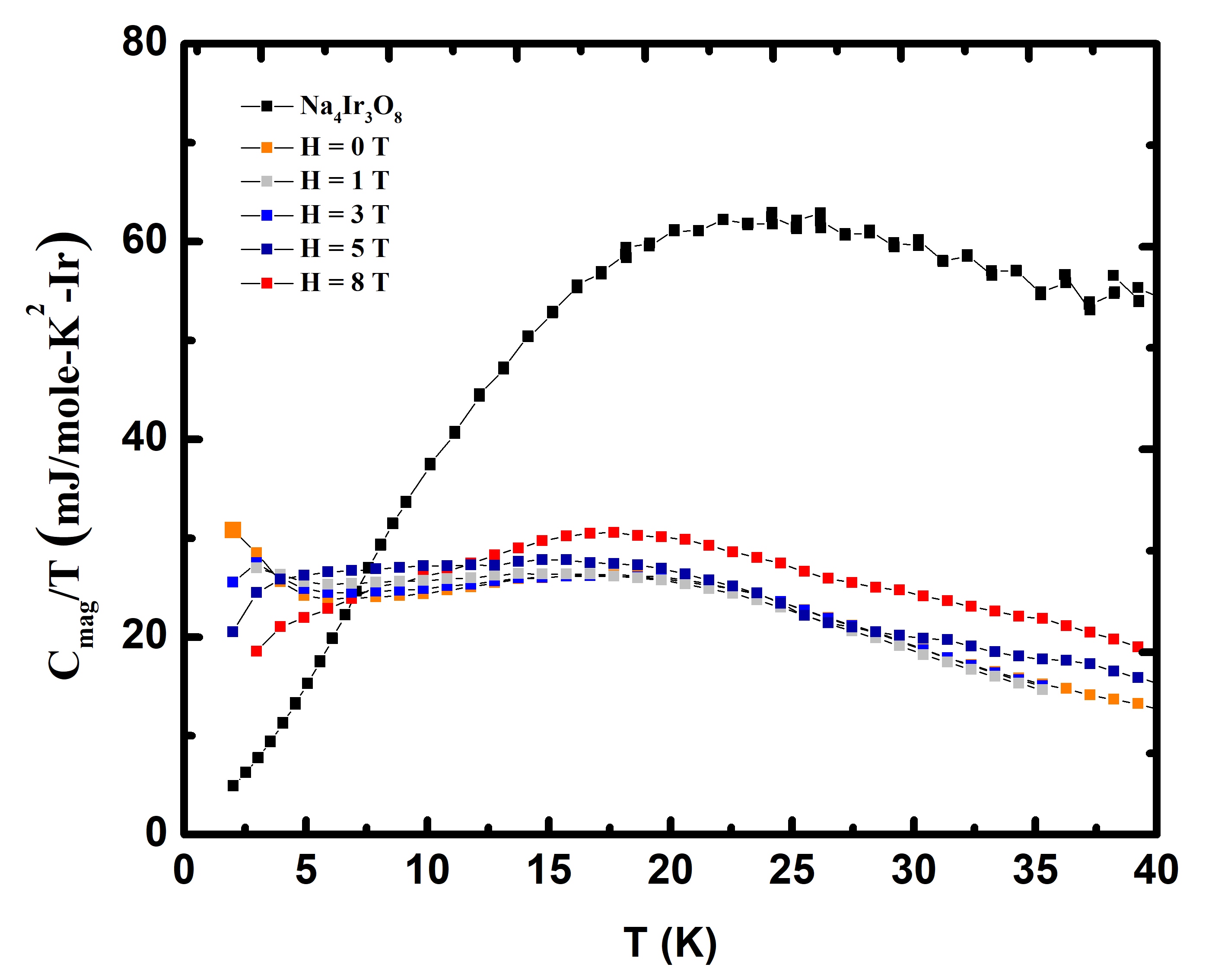}
	\caption{ (a) Heat capacity C$_{p}$ versus Temperature T(K) for Na$_4$Ir$_{2.7}$Ti$_{0.3}$O$_8$ at different applied magnetic fields up to 9 T. Inset shows the low-temperature variation of C$_{p}$/T for Na$_4$Ir$_{2.7}$Ti$_{0.3}$O$_8$ at different applied magnetic fields. (b) C$_{mag}$/T versus T for Na$_4$Ir$_{2.7}$Ti$_{0.3}$O$_8$).} 
 		\label{hc2}
 	\end{figure}

The $C_{mag}$ data for the Ti substituted Na$_4$Ir$_{2.7}$Ti$_{0.3}$O$_8$ measured at various magnetic fields are shown in Fig.~\ref{hc2}. For comparision the $C_{mag}$ data for Na$_4$Ir$_{3}$O$_8$ is also plotted.  The broad anomaly seen in $C_{mag}$ for Na$_4$Ir$_{3}$O$_8$ is completely suppressed for the Ti substituted material.  The $C_{mag}$ also shows a strong magnetic field dependence.
 
 \section{Conclusion}
The primary motive of this study is to understand the bond disorderd spin-liquid state. By substitution of Ru/Ti for Ir, we disturbed the magnetic interactions between the Ir atoms in Na$_4$Ir$_3$O$_8$. 
Specifically, we have synthesized polycrystalline materials Na$_4$(Ir$_{1-x}$Ru$_x$)$_3$O$_8~ (x = 0, 0.05, 0.1, 0.2)$ and Na$_4$Ir$_{2.7}$Ti$_{0.3}$O$_8$ and studied their structural, magnetic, and thermal properties.  
Our PXRD results confirm successful substitution of Ru/Ti for Ir in Na$_4$Ir$_3$O$_8$. Both Ru and Ti have a ionic radii smaller than Ir which leads to a decreases in the unit cell size. A rietveld refinement also indicates slight Na vacancy for all substituted materials.  

The main features of interest in the parent material Na$_4$Ir$_3$O$_8$ are it's Mott insulating state, a large Weiss temperature $\theta \approx -650$~K, no long ranged ordering down to very low temperatures, a pronounced anomaly in the magnetic heat capacity $C_{mag}$ peaked around $25$~K, power-law $C \sim T^\alpha$ with $\alpha \approx 2.6$, and a spin-freezing below $T_f \approx 6$~K ~\cite{okamato, yogesh}.  Our recent work has already shown that even large Na vacancies have weak effect on these features of the parent compound \cite{ashiwini}.  Therefore, we can rule out the changes we see in the current samples as coming from Na vacancies.

We find that the Mott insulating state persists in the Ru/Ti substituted materials.  For Na$_4$(Ir$_{1-x}$Ru$_{x}$)$_3$O$_8$, the Ru substitution leads to an increase in the average effective moment, suggesting that Ru dopants go in as localized moments.  The Weiss temperature stays antiferromagnetic and increases with $x$.  This is most likely due to an enhancement of magnetic exchange due to compression of the lattice.  The spin-freezing temperature $T_f$ increases with $x$ and reaches $T_f \approx 12.4$~K for $x = 0.2$.  This reflects the increasing disorder in the partially substituted samples which leads to an enhanced freezing temperature.  This also suggests that the parent material is inherently disordered and efforts to synthesize less disordered Na$_4$Ir$_3$O$_8$ should be made to reveal the true ground state which might well be the much sought after $3$-dimensional quantum spin liquid. 

The anomaly in $C_{mag}$ is progressively suppressed with increasing $x$ suggesting that Ru substitution disrupts the magnetic sublattice and disturbs the mechanism (short ranged order!) leading to the anomaly.  The $C$ data continues to show a magnetic field insensitive power-law behaviour $C \sim T^\alpha$ for all $x$ and the exponent $\alpha$ falls towards $2$ with increasing $x$.   

For the non-magnetic Ti substituted material Na$_4$Ir$_{2.7}$Ti$_{0.3}$O$_8$ we find different behaviours.  The magnetic correlations weaken as evidenced by a smaller Weiss temperature, the spin-freezing is pushed to much lower temperature $T_f \approx 2.3$~K suggesting a percolation mechanism, and the anomaly in $C_{mag}$ is completely suppressed.

Since both Ru and Ti are smaller than Ir, their partial substitution for Ir in Na$_4$Ir$_3$O$_8$ leads to positive chemical pressure.  Therefore, the contrasting effects on the magnetic properties of Na$_4$Ir$_3$O$_8$ must be due to the kind of bond-disorder (magnetic or non-magnetic) introduced by their substitution.  Further studies like NMR and ${\mu}$SR on Ru/Ti substituted materials can give important microscopic insights to help our understanding of Na$_4$Ir$_3$O$_8$.

\end{document}